\documentclass[conference]{IEEEtran}

\usepackage[utf8]{inputenc}  
\usepackage{amsmath, amssymb} 
\usepackage{graphicx}        
\usepackage{cite}            
\usepackage{url}             
\usepackage{algorithm}     
\usepackage{algorithmic}   
\usepackage{float} 
\usepackage{multirow}
\usepackage[font=footnotesize,labelfont=bf,labelsep=period]{caption}
\begin{document}

\title{Lightweight MobileNetV1+GRU for ECG Biometric Authentication: Federated and Adversarial Evaluation}

\author{\IEEEauthorblockN{Dilli Hang Rai}
\IEEEauthorblockA{Department of Computer Science \& Information Technology\\
IoST, Tribhuwan University, Nepal\\
dillihangrai.078@godawari.edu.np}
\and
\IEEEauthorblockN{Sabin Kafley}
\IEEEauthorblockA{Department of Electronics and Computer Engineering\\
Institute of Engineering, Dharan, Nepal\\
sabinkafley@tu.edu.np}
}

\maketitle

\begin{abstract}
ECG biometrics offer a unique, secure authentication method, yet their deployment on wearable devices faces real-time processing, privacy, and spoofing vulnerability challenges. This paper proposes a lightweight deep learning model (MobileNetV1+GRU) for ECG-based authentication, injection of 20dB Gaussian noise \& custom preprocessing. We simulate wearable conditions and edge deployment using the ECGID, MIT-BIH, CYBHi, and PTB datasets, achieving accuracies of 99.34\%, 99.31\%, 91.74\%, and 98.49\%, F1-scores of 0.9869, 0.9923, 0.9125, and 0.9771, Precision of 0.9866, 0.9924, 0.9180 and 0.9845, Recall of 0.9878, 0.9923, 0.9129, and 0.9756, equal error rates (EER) of 0.0009, 0.00013, 0.0091, and 0.0009, and ROC-AUC values of 0.9999, 0.9999, 0.9985, and 0.9998, while under FGSM adversarial attacks, accuracy drops from 96.82\% to as low as 0.80\%. This paper highlights federated learning, adversarial testing, and the need for diverse wearable physiological datasets to ensure secure and scalable biometrics. 
\end{abstract}


\section{Introduction}
Electrocardiogram (ECG)-based biometrics provide secure authentication for wearables due to their uniqueness and resistance to replication. ECG waveforms vary across individuals, making them a secure biometric \cite{Biel2001}. Traditional ML struggles with temporal ECG patterns, while conventional traits like fingerprints or face can be spoofed or altered \cite{AlJibreen2024,Beritelli2007,Camara2015}.  Despite advances in ECG-based biometrics, gaps remain in lightweight models that handle noisy signals, support cross-dataset evaluation, preserve privacy, and resist adversarial attacks.  
Centralized processing risks privacy, while federated learning keeps data local \cite{Celik2025}; however, both remain vulnerable to adversarial attacks. Previous ECG authentication studies face key limitations: small datasets and hardware dependence \cite{Kim2020}, lower accuracy on long windows \cite{BicakciYesilkaya2025}, poor cross-session/dataset generalization \cite{Ammour2023}, and issues with controlled acquisition, noise, and limited diversity \cite{Prakash2023,Hazratifard2023,Xu2024}. Real-world use remains hindered by FAR/FRR imbalance and scalability challenges \cite{AlJibreen2024}. We improved ECG authentication by using short R-peak windows (0.6s, 4s) \cite{BicakciYesilkaya2025}, MobileNetV1 with CWT-based R-peak detection \cite{MalekiLonbar2024}, and multiple datasets (ECG-ID, PTB, MIT-BIH, CYBHI) for better generalization. Robustness was enhanced with noise injection \cite{Prakash2023} and fusion-based temporal modeling \cite{Ammour2023}. Security and scalability were evaluated through FGSM and FedAvg \cite{Xu2024}. The contributions of this paper can be defined as first to fully configure \& train the lightweight MobileNetV1+GRU ECG authentication framework, achieving the impressive accuracies of 99.34\% and 96.82\%  before and after attack.
Hybrid preprocessing, ECG signals were injected with 20dB Gaussian noise, R-peaks were centered via custom beat segmentation for reliable templates, and signals were downsampled to simulate wearable conditions. Cross-Dataset Evaluation, the proposed architecture was trained, validated, and tested on all four datasets—ECG-ID, MIT-BIH, CYBHi, and PTB—with variety of subjects. Integration of federated learning with FGSM attack evaluation to ensure robustness, along with Explainable AI (XAI) for model interpretability.

This paper is organized as follows: Section II presents the methodology, detailing the proposed approach. Section III describes the experimental setup, including tables, and Section IV explains the results, and analyses with corresponding figures and diagrams. Finally, Section VI concludes the study.

\section{Methodology}
The proposed model combines MobileNetV1 for spatial feature extraction and This section describes dataset collection, preprocessing, transformation, and the proposed model.

\subsection{Dataset}
Four ECG datasets were used: ECG-ID (90 subjects, lead I, 500Hz), MIT-BIH (47 subjects, 2 leads, 360Hz), PTB (290 subjects, 15 leads, 1000Hz), and CYBHi (125 subjects, 1000Hz, multiple sites), split 70:15:15 for training, validation, and testing~\cite{goldberger2000physionet,mark1982annotated,bousseljot1995cardiodat,pereira2023biometric,daSilva2013cybhi}.
\vspace{-0.3cm}

\begin{figure}[h!]
\centering
\includegraphics[width=0.5\textwidth]{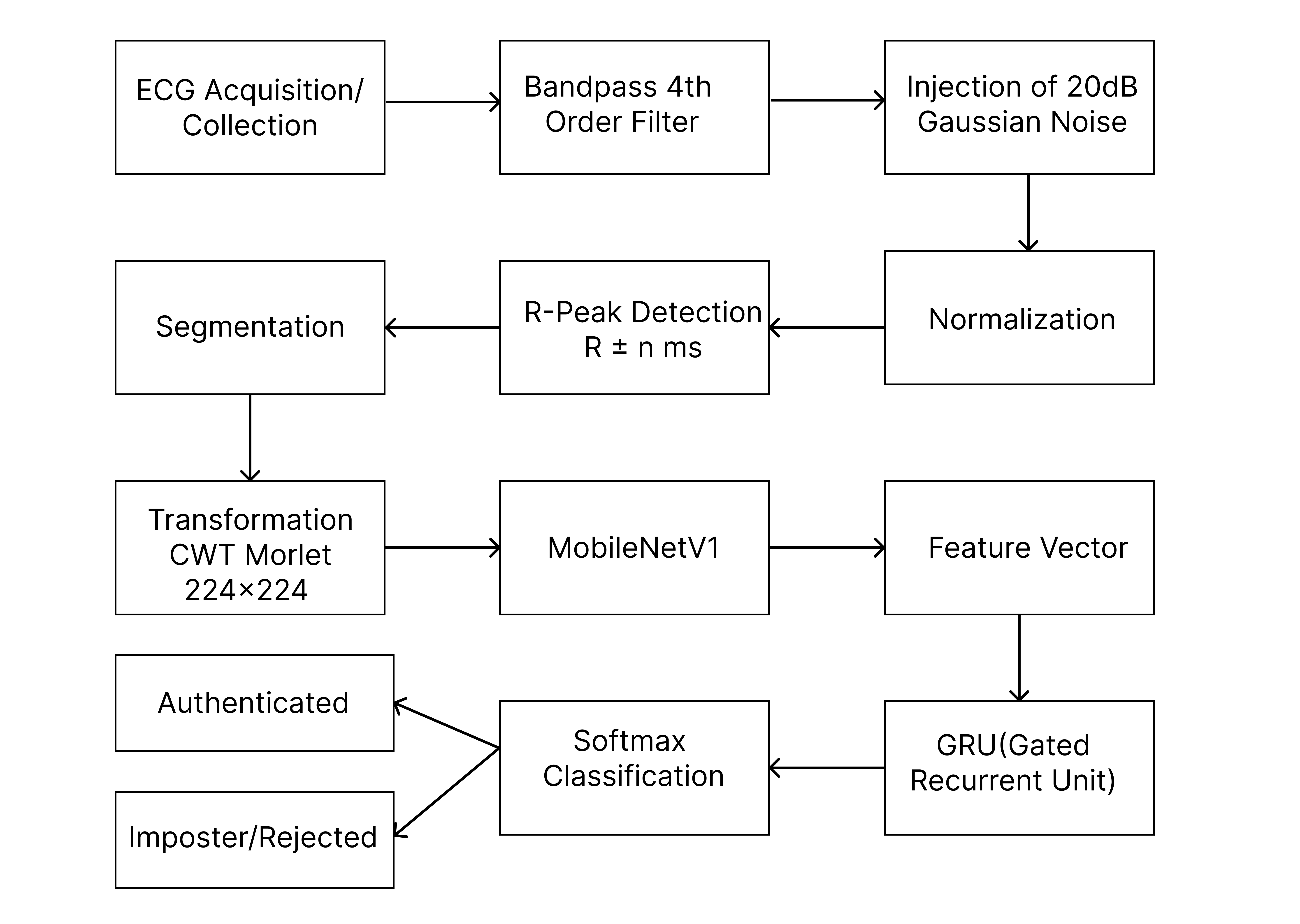}
\caption{Proposed MobileNetV1+GRU architecture for ECG authentication.}
\end{figure}

\subsection{Data Preprocessing}
\subsubsection{Bandpass Filtering,Noise Injection,Normalization,R-Peak Detection}

A 4\textsuperscript{th}-order bandpass filter (0.5--40\,Hz) removed baseline and high-frequency noise, followed by 20\,dB Gaussian noise injection and Z-score normalization to simulate wearable conditions. 
\vspace{-0.3cm}
\subsubsection{R-Peak Detection and Segmentation}
\begin{figure}[h!]
\centering
\includegraphics[width=0.6\linewidth]{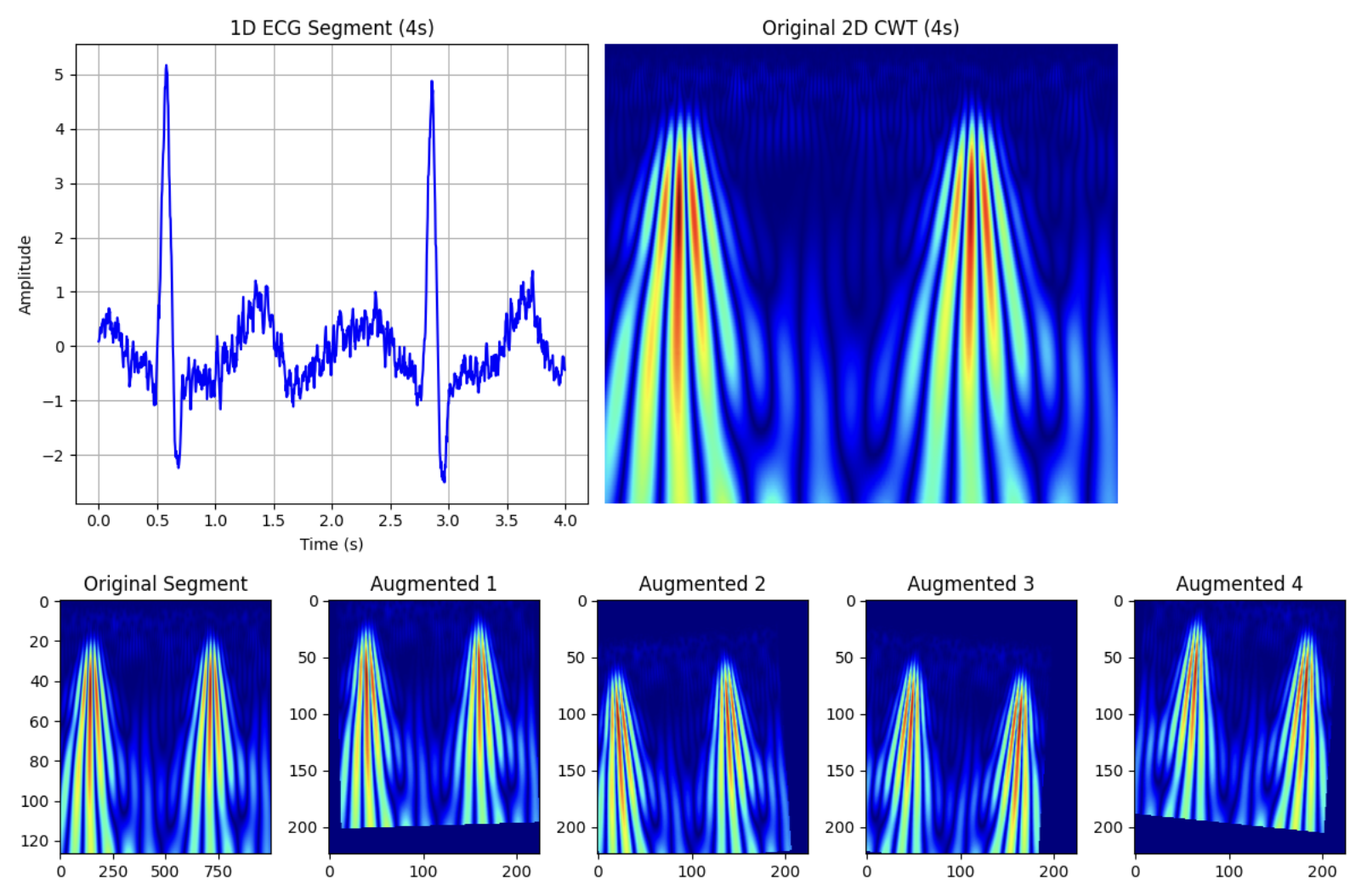}
\caption{(a) R-peak segmentation for 4-second CYBHi with transformation \& augmentation}
\label{fig:cybhi_a}
\end{figure}

Detecting R-peaks facilitates segmenting heartbeats, identifying and passing the local temporal context features to the model, and centering the R-peak with $R \pm n$ ms, where $n$ is a constant value. We used the \texttt{find\_peaks()} function from SciPy instead of the Pan-Tompkins algorithm (Fig.~\ref{fig:cybhi_a}) due to its lower computational complexity.
With a sampling frequency of $F_s = 250$ Hz, each sample corresponds to 4 ms, and for a 500 ms window before and after the R-peak, the number of samples is $500~\text{ms} / 4~\text{ms/sample} = 125$ samples; thus, the total segment length around the R-peak is 125 samples before the R-peak, 1 sample at the R-peak, and 125 samples after, giving a total of 251 samples, and the segmented beat can be written as $b[n] = x[R - 125 : R + 125]$.

\begin{figure}[h!]
\centering
\includegraphics[width=0.6\linewidth]{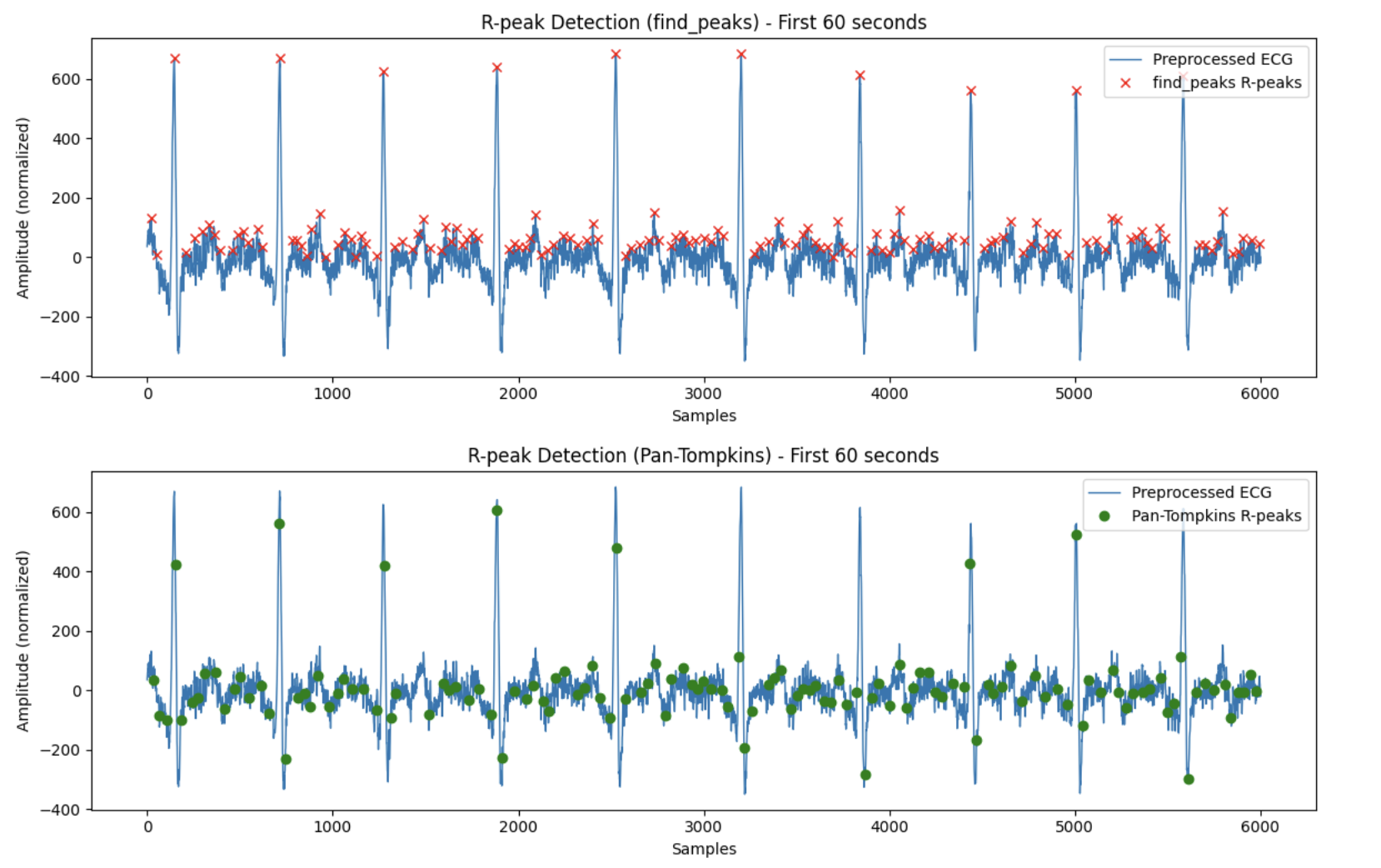}
\caption{(b) R-peak detection using \texttt{find\_peaks()} and Pan-Tompkins algorithm}
\label{fig:cybhi_b}
\end{figure}

\begin{figure}[h!]
\centering
\includegraphics[width=0.5\linewidth]{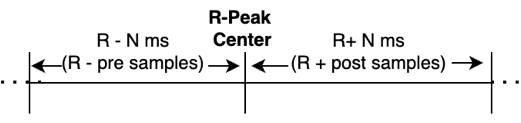}
\caption{(c) Equation for R-peak centered Segment}
\label{fig:cybhi_c}
\end{figure}

\subsubsection{Transformation and Augmentation}
Each ECG beat was converted to a $224 \times 224$ CWT--Morlet scalogram. Training images were augmented (rotation $\pm5^\circ$, 10\% translation, scaling 0.9--1.1, horizontal flip), while validation and test sets were resized and ImageNet-normalized; no augmentation was applied.

\subsection{Proposed Hybrid Model MobileNetV1 + GRU}
In our proposed hybrid architecture, the MobileNetV1 model extracts 1280-dimensional spatial and spectral features from $224 \times 224$ ECG scalograms, GRU repeats them into a sequence of length 49 (default), and processes them with a GRU of 128 hidden units (default). The final hidden state passes through an optional dropout, a 128-unit fully connected layer with activation, and a final fully connected layer to output class probabilities (softmax classification).

\label{sssec:proposed_hybrid_mobilenetv1+gru}
\begin{algorithm}[H]
\caption{MobileNetV1-GRU Hybrid ECG Biometric Authentication}
\begin{enumerate}
  \item \textbf{Input:} Batch of ECG scalogram images $X \in \mathbb{R}^{B \times 3 \times 224 \times 224}$; 
        \textbf{Output:} Class predictions $\hat{Y} \in \mathbb{R}^{B \times C}$

  \item Extract spatial features using MobileNetV1:
  \begin{equation*}
  F \gets \text{MobileNetV1}(X), \quad F \in \mathbb{R}^{B \times 1280}
  \end{equation*}

  \item Add a sequence dimension:
  \begin{equation*}
  S \gets \text{reshape}(F, (B, 1, 1280))
  \end{equation*}

  \item Repeat feature vector to create an artificial sequence of length 49:
  \begin{equation*}
  S \gets \text{repeat}(S, (1, 49, 1)), \quad S \in \mathbb{R}^{B \times 49 \times 1280}
  \end{equation*}

  \item Process the sequence using GRU:
  \begin{equation*}
  H, _ \gets \text{GRU}(S)
  \end{equation*}

  \item Extract the final hidden state of the GRU:
  \begin{equation*}
  H \gets H[:, -1, :] , \quad H \in \mathbb{R}^{B \times \text{GRU\_units}}
  \end{equation*}

  \item Apply dropout for regularization (optional):
  \begin{equation*}
  Y \gets \text{Dropout}(H, p)
  \end{equation*}

  \item Pass through fully connected layers with activation:
  \begin{equation*}
  Y \gets \text{Activation}(W_n Y + b_n), \quad n = 1,2,\dots
  \end{equation*}

  \item Compute final class probabilities via Softmax:
  \begin{equation*}
  \hat{Y} \gets \text{Softmax}(W_{\text{final}} Y + b_{\text{final}}), \quad \hat{Y} \in \mathbb{R}^{B \times C}
  \end{equation*}

\end{enumerate}
\end{algorithm}

\newpage{}
\section{Experimental Setup}
In this section, we describe the dataset, preprocessing steps, experimental design, evaluation metrics, and implementation details used to validate the proposed methodology.

As summarized in Table~\ref{tab:PreprocessingHyperparameters}, the proposed architecture was shaped after rigorous training, validation, and systematic hyperparameter tuning. A 0.6-second R-peak window captures a single heartbeat (assuming 100bpm), while a 4-second window includes multiple beats, increasing spatial context. Leaky ReLU is applied to address dying ReLU in large PTB and CYBHi softmax-class datasets.

\begin{table}[h!]
\centering
\caption{{Preprocessing \& Hyperparameter Settings}}
\label{tab:PreprocessingHyperparameters}
\footnotesize
\begin{tabular}{|p{2cm}|p{1cm}|p{1cm}|p{1.25cm}|p{1.25cm}|}
\hline
\textbf{Parameter} & \textbf{ECG-ID} & \textbf{MIT-BIH} & \textbf{CYBHI} & \textbf{PTB} \\
\hline
Train Samples & 20613 & 50400 & 13115 & 32844 \\
\hline
Validation Samples & 4384 & 10800 & 1804 & 4016 \\
\hline
Test Samples & 4507 & 10800 & 1924 & 4382 \\
\hline
Downsampled Hertz & 100 Hz & 100 Hz & 250 Hz & 250 Hz \\
\hline
Min Distance between RR-interval & 0.6 sec & 0.6 sec & 4 sec & 4 sec \\
\hline
Min Height Threshold & 0.5 & 0.5 & 0.5 & 0.5 \\
\hline
Epochs & 100 & 50 & 80 & 100 \\
\hline
Learning Rate & 0.0001 & 0.0001 & 0.0001 & 0.001 \\
\hline
Dropout & 0.35 & 0.35 & 0.00 & 0.00 \\
\hline
Activation Function Between GRU Layers & ReLU & ReLU & LeakyReLU & LeakyReLU \\
\hline
Regularization & L2 & GRU Layer 1 with L2:0.01 & GRU Layer 1 with L2:0.01 & - \\
\hline
GRU Layer 1 Units & 128 & 128 & 64 & 128 \\
\hline
Feature Extractor (MobileNetV1 CNN layers), fc1, fc2 with L2 Between Layers & 0.01 & 0.01 & - & - \\
\hline
Data Augmentation & Yes & Yes & Yes & Yes \\
\hline
Batch Size & 64 & 64 & 8 & 16 \\
\hline
LR Scheduler: Patience & 5 & 5 & 5 & 8 \\
\hline
LR Scheduler: Factor & 0.1 & 0.1 & 0.1 & 0.1 \\
\hline
LR Scheduler: Min-LR & 1e-8 & 1e-8 & 1e-6 & 1e-6 \\
\hline
Dataset Balancing Technique & Class Weights & Manually Balanced & Class Weights & Class Weights \\
\hline
Optimizer & Adam & Adam & Adam & Adam \\
\hline
CWT & Morlet & Morlet & Morlet & Morlet \\
\hline
Peak Detection & R-Peak ± Segment & R-Peak, R±250ms & R-Peak, R±250ms & R-Peak, R±500ms \\
\hline
\end{tabular}
\end{table}

\section{Results and Discussion}
In this section, the experimental setup and results are presented in figures and summarized in tables.

\begin{figure}[htb]
\centering

\begin{minipage}[b]{0.48\linewidth}
  \centering
  \includegraphics[width=\linewidth]{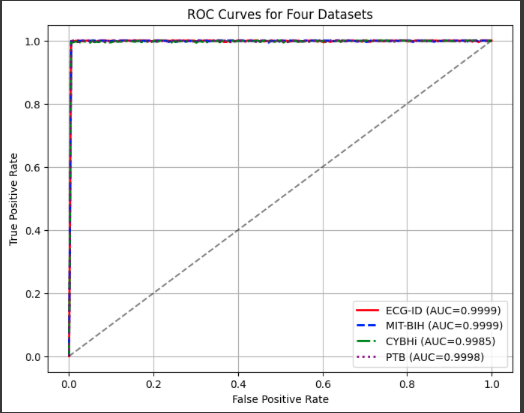}
  \centerline{(a) EER Curves }
\end{minipage}
\hfill
\begin{minipage}[b]{0.48\linewidth}
  \centering
  \includegraphics[width=\linewidth]{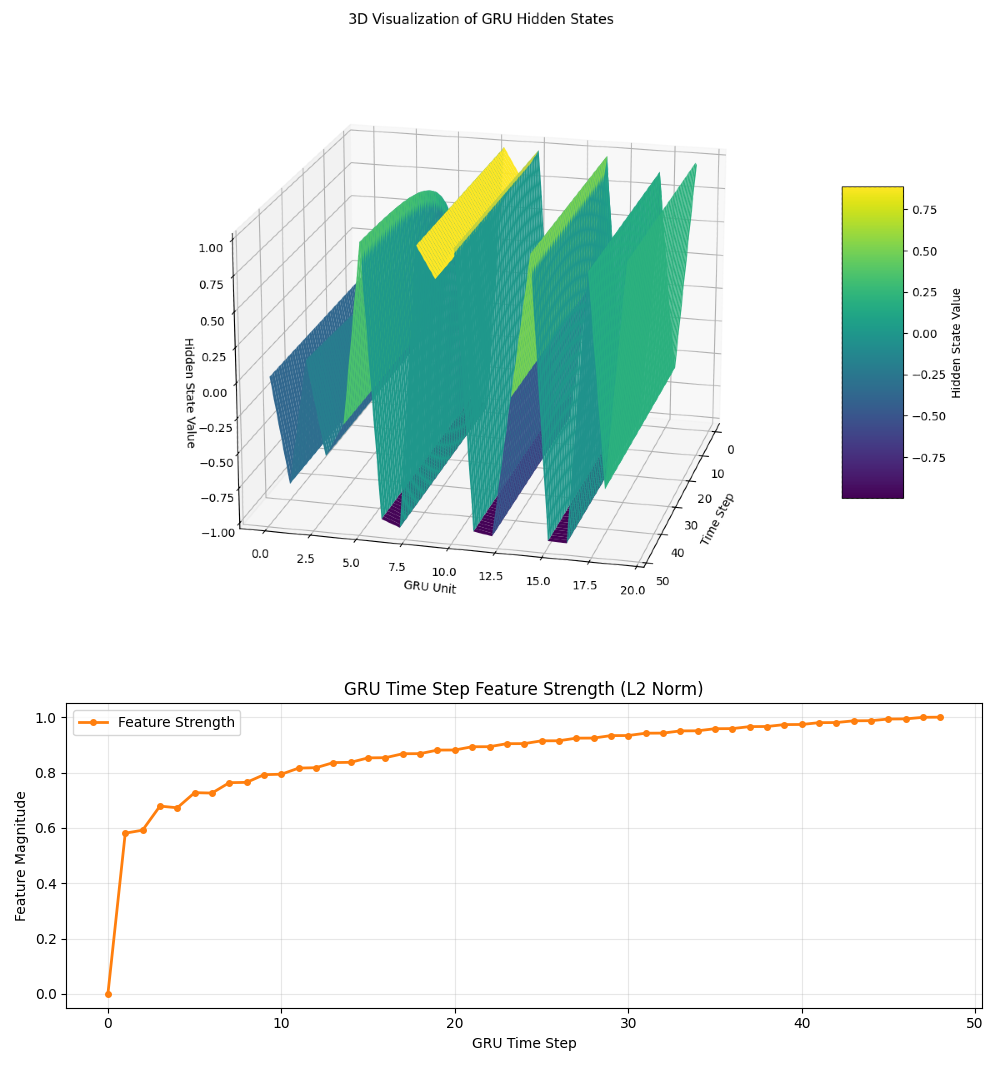}
  \centerline{(b) 3D GRU Feature Strength}
\end{minipage}

\vspace{0.4cm} 

\begin{minipage}[b]{0.8\linewidth}
  \centering
  \includegraphics[width=\linewidth]{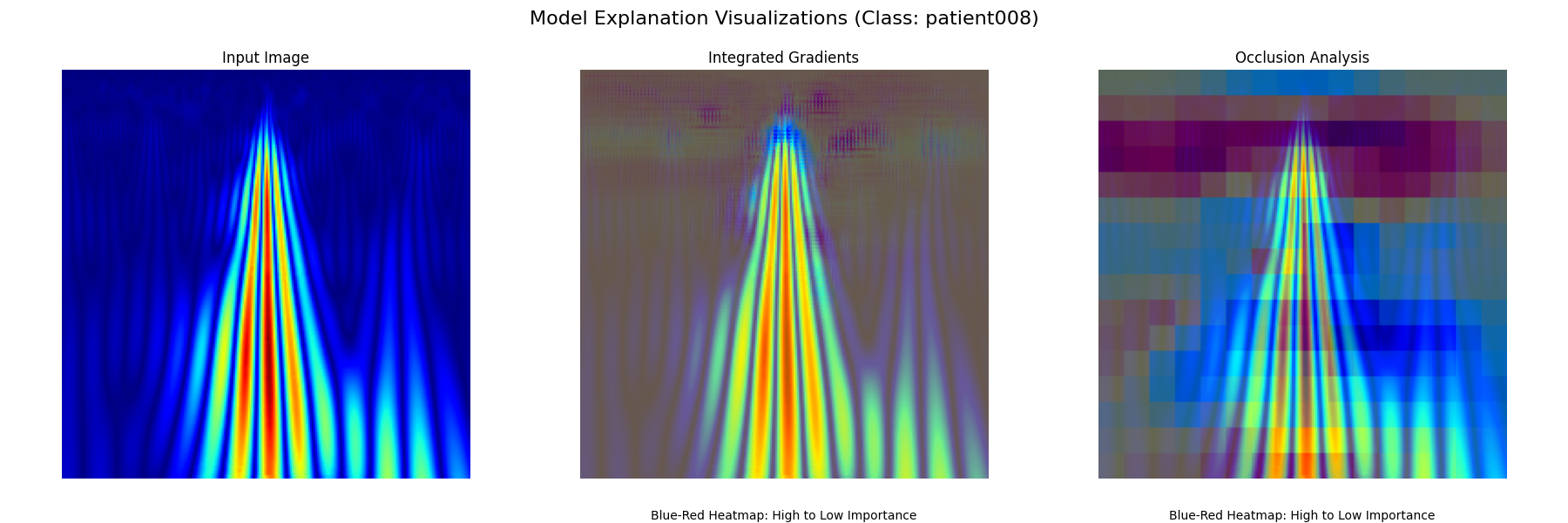}
  \centerline{(c) PTB008 Occlusion-based IG Highlighting Model Decisions}
\end{minipage}

\caption{Explainable AI (XAI) results: (a) EER curves of Model Trained with Four Datasets, (b) 3D GRU feature strength, and (c) PTB008 occlusion-based IG.}
\label{fig:xai_results}
\end{figure}

\begin{figure}[htb]
\centering

\begin{minipage}[b]{0.45\linewidth}
  \centering
  \includegraphics[width=\linewidth]{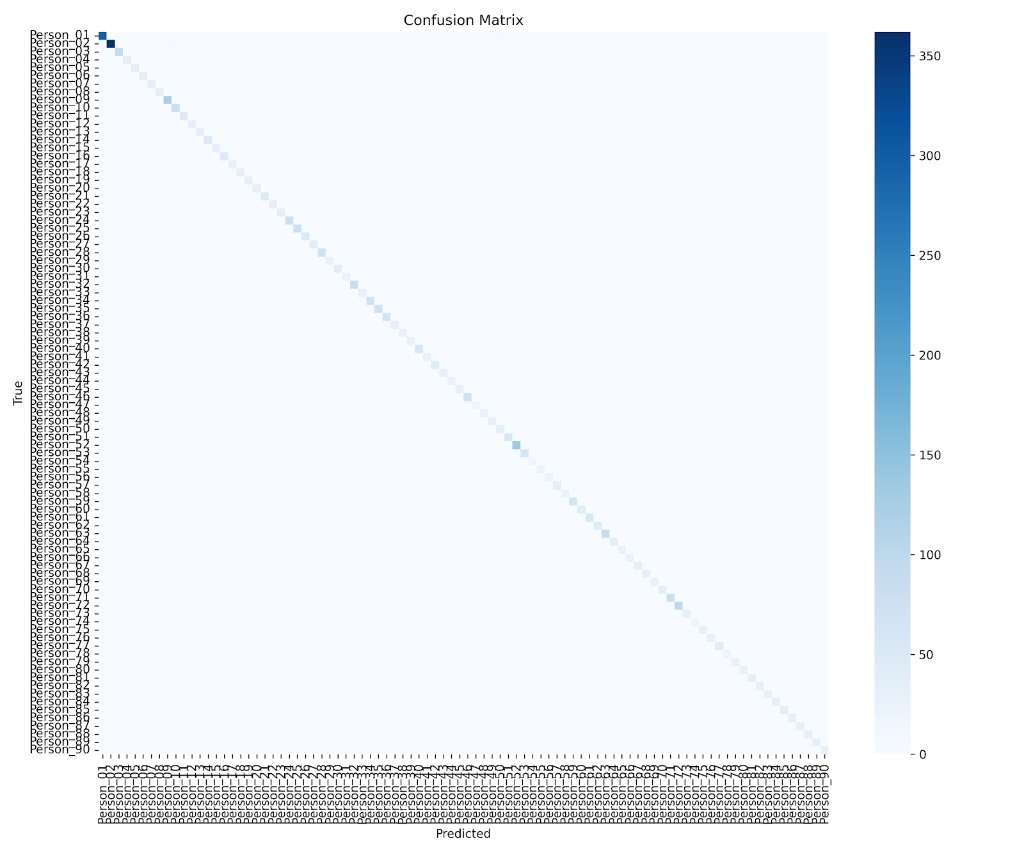}
  \centerline{(a) ECG-ID Confusion Matrix}
\end{minipage}
\hfill
\begin{minipage}[b]{0.45\linewidth}
  \centering
  \includegraphics[width=\linewidth]{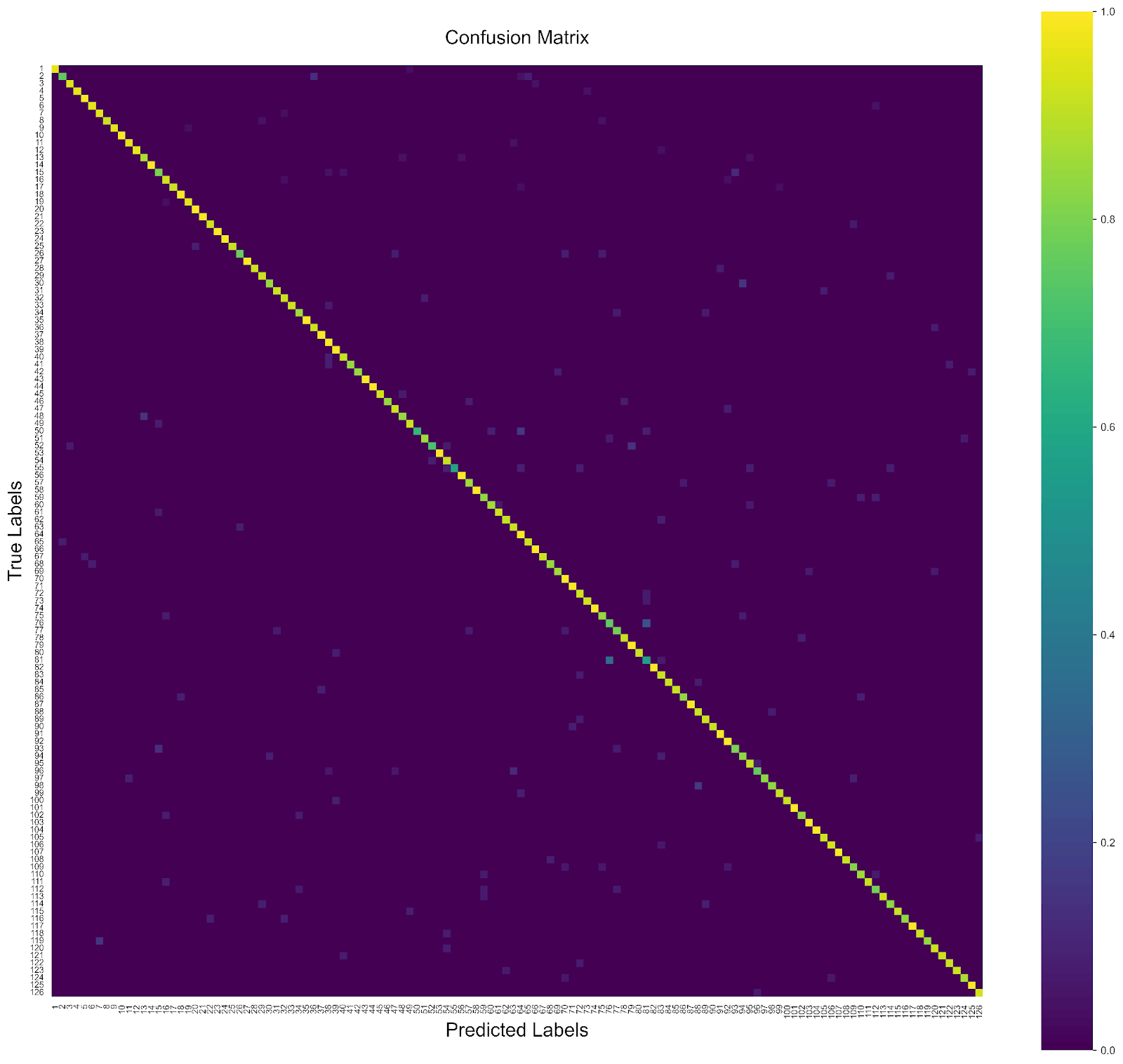}
  \centerline{(b) CYBHi Confusion Matrix}
\end{minipage}

\vspace{0.4cm} 

\begin{minipage}[b]{0.45\linewidth}
  \centering
  \includegraphics[width=\linewidth]{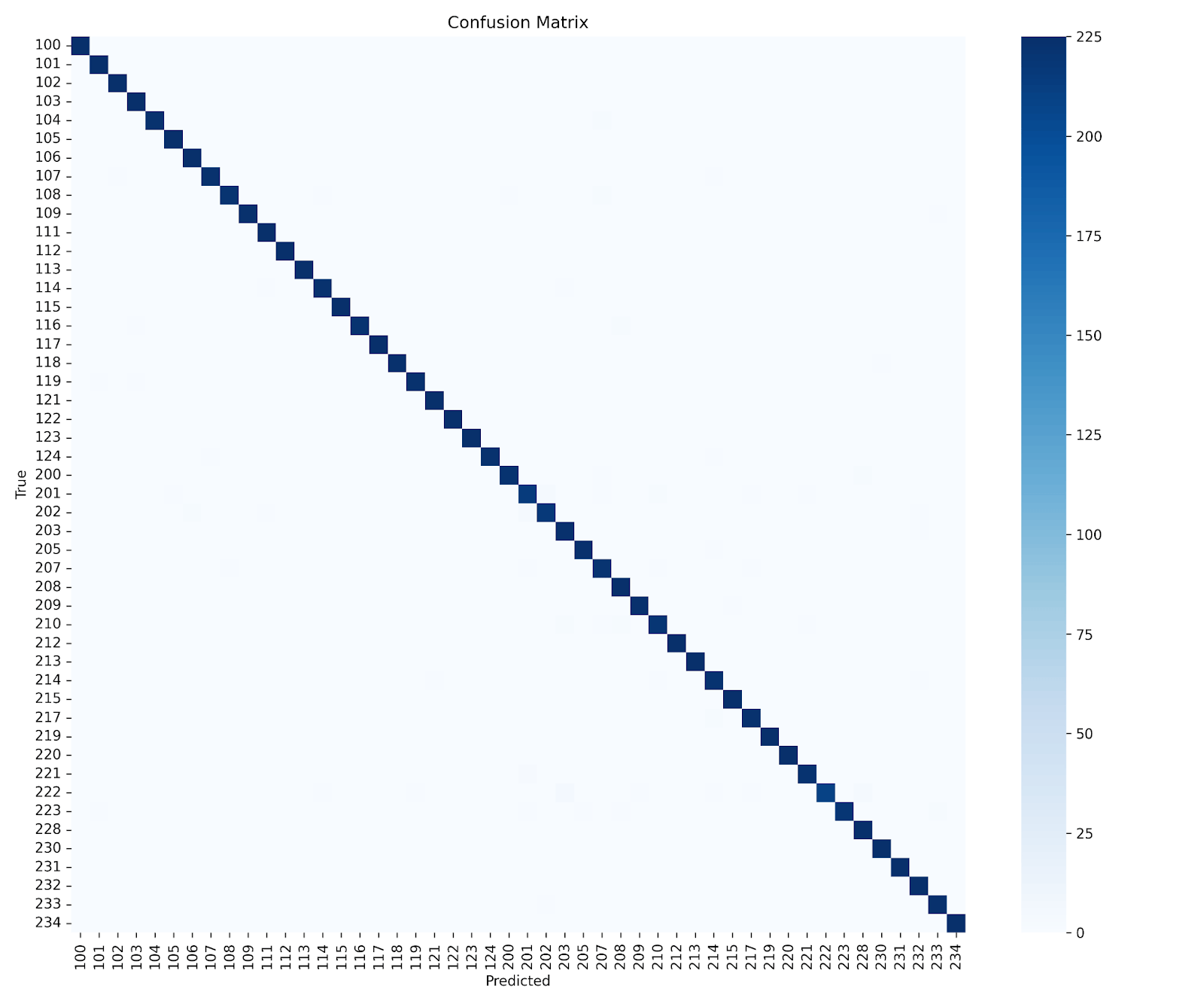}
  \centerline{(c) MIT-BIH Confusion Matrix}
\end{minipage}
\hfill
\begin{minipage}[b]{0.45\linewidth}
  \centering
  \includegraphics[width=\linewidth]{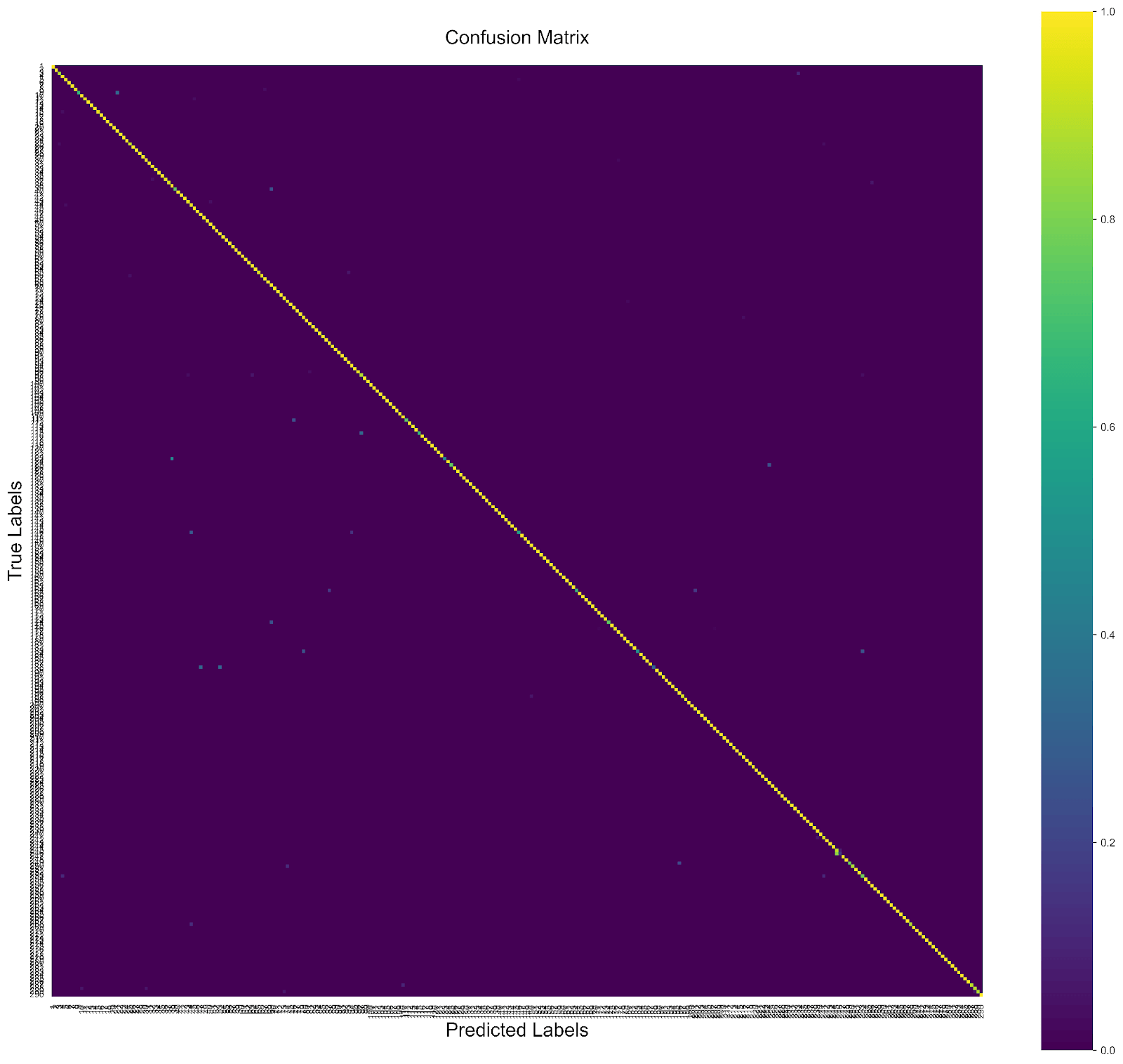}
  \centerline{(d) PTB Confusion Matrix}
\end{minipage}

\caption{Confusion Matrices For Four Datasets: (a) ECG-ID, (b) CYBHi, (c) MIT-BIH, and (d) PTB.}
\label{fig:confusion_matrices}
\end{figure}

\begin{figure}[htb]
\centering
\includegraphics[width=\linewidth]{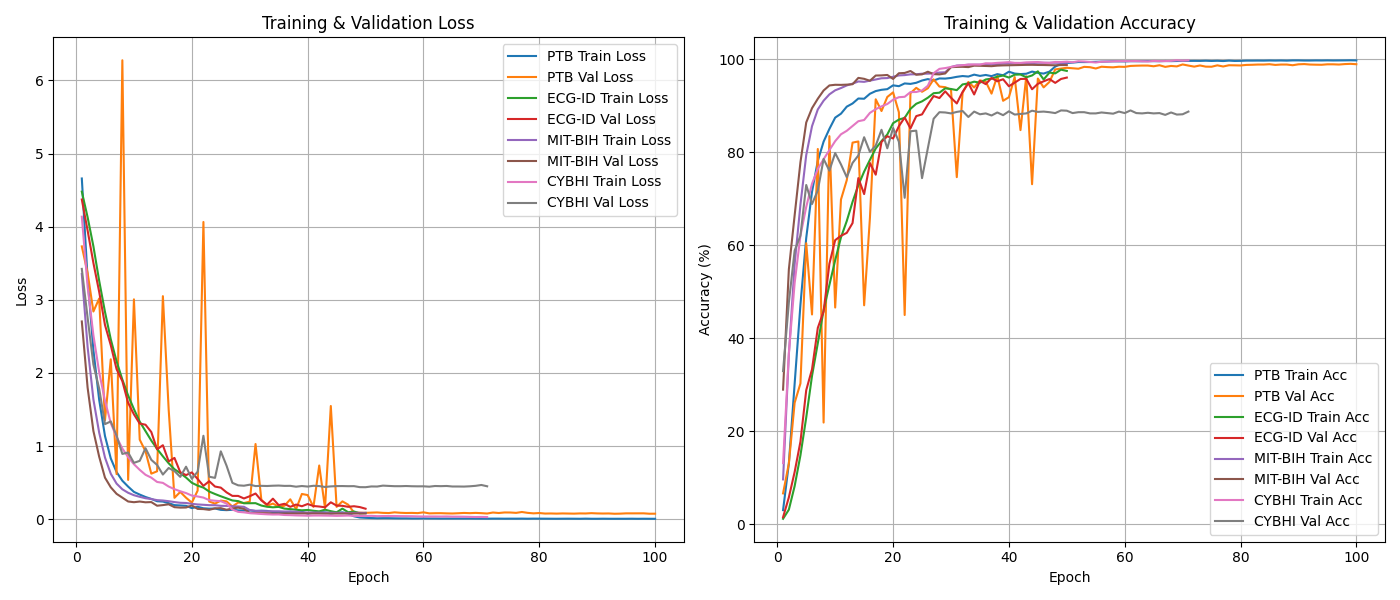}
\caption{Model Training History with Four Datasets (ECG-ID, MIT-BIH, CYBHi, and PTB).}
\label{fig:training_history}
\end{figure}

\begin{table}[htbp]
\centering
\caption{{Effect of FGSM Adversarial Attacks on Proposed Trained ECG Biometric Hybrid Model with Datasets}}
\label{tab: FGSM Adversarial Attacks Results Transposed}
\footnotesize
\begin{tabular}{|p{1.85cm}|p{1.15cm}|p{1.15cm}|p{1.15cm}|p{1cm}|}
\hline
\textbf{} & \textbf{ECG-ID} & \textbf{MIT-BIH} & \textbf{CYBHi} & \textbf{PTB} \\
\hline
Base Acc (20 dB) & 99.31\% & 99.34\% & 91.74\% & 99.22\% \\
\hline
Acc $\epsilon$=0.0001 & 89.19\% & 96.82\% & 29.83\% & 70.04\% \\
\hline
Acc $\epsilon$=0.001 & 57.49\% & 88.13\% & 20.63\% & 63.33\% \\
\hline
Acc $\epsilon$=0.1 & 0.80\% & 0.87\% & 17.57\% & 9.77\% \\
\hline
Loss $\epsilon$=0.0001 & 034 & 0.1377 & 3.5475 & 2.1662 \\
\hline
Loss $\epsilon$=0.001 & 2.3652 & 787 & 5.2143 & 4.7083 \\
\hline
Loss $\epsilon$=0.1 & 11.2910 & 6.5079 & 6.2293 & 22.9676 \\
\hline
\end{tabular}
\end{table}

\begin{table}[h!]
\centering
\caption{{Federated Learning: Client Accuracies, Expected Accuracies, and Global Model Update Summary}}
\label{tab: FedAvgSummary}
\footnotesize
\begin{tabular}{|p{1.5cm}|p{1.15cm}|p{1.15cm}|p{1.15cm}|p{1.15cm}|p{1.5cm}|}
\hline
\textbf{Client / Model} & \textbf{Acc.} & \textbf{Weight Sum} & \textbf{Expected Acc.} & \textbf{Remarks} \\
\hline
Client 1 & 97.20\% & – & $\sim$95\% & Impressive \\
\hline
Client 2 & 98.90\% & – & $\sim$95\% & Impressive \\
\hline
Client 3 & 98.79\% & – & $\sim$95\% & Impressive\\
\hline
Global Before Update & – & 0.3845 & – & - \\
\hline
Global After Update (Round 1) & 23.47\% & 0.5795 & $\sim$25\% & Updated FedAvg  \\
\hline
\end{tabular}
\end{table}

FedAvg demonstrates effective aggregation and a promising approach for robust, privacy-preserving frameworks, as shown in Table~\ref{tab: FedAvgSummary}. Although a naive method, we highlight and assess the lightweight model, simulating wearable like scenarios. White-box FGSM attacks showed models trained on ECG-ID, PTB, and MIT-BIH were highly robust, while CYBHi—collected under less controlled conditions—performed worse, emphasizing that dataset quality and training approach critically impact robustness (Table~\ref{tab: FGSM Adversarial Attacks Results Transposed}).

\begin{table}[h!]
\centering
\caption{ Model Final Performance Metrics Across ECG Datasets}
\label{tab:results}
\footnotesize
\begin{tabular}{|p{2.25cm}|p{1cm}|p{1.25cm}|p{1cm}|p{1cm}|}
\hline
\textbf{Metric} & \textbf{ECGID} & \textbf{MIT-BIH} & \textbf{CYBHI} & \textbf{PTB} \\
\hline
Subjects & 90 & 48 & 216 & 290 \\
\hline
Noise DB & 20 & 20 & 20 & 20 \\
\hline
Test Acc & 99.34\% & 99.31\% & 91.74\% & 98.49\% \\
\hline
Test Loss & 0.0506 & 0.0258 & 0.3330 & 0.0992 \\
\hline
Precision (Macro) & 0.9866 & 0.9924 & 0.9180 & 0.9845 \\
\hline
Top-5 Acc & 99.82\% & 99.82\% & 97.87\% & 99.75\% \\
\hline
F1-score (Macro) & 0.9869 & 0.9923 & 0.9125 & 0.9771 \\
\hline
Recall (Macro) & 0.9878 & 0.9923 & 0.9129 & 0.9756 \\
\hline
EER (Macro-average) & 0.0009 & 0.0046 & 0.0895 & 0.0009 \\
\hline
EER (Top-5) & 0.0013 & 0.0091 & 0.0169 & 0.0026 \\
\hline
ROC AUC (Macro) & 0.9999 & 0.9999 & 0.9985 & 0.9998 \\
\hline
\end{tabular}
\end{table}

The Table~\ref{tab:results}, ~\ref{tab:ecg_comparison} shows that our method achieved outperformed existing methods with ECG-ID 99.34\% the highest accuracy. While CYBHi collects ECG via finger or palm contact using wearables, making it inherently noisy, which could be the reason for its lowest accuracy, as CNN activations on its signals are more variable and wide-ranging compared to MIT datasets~\cite{Silva2020,Silva2019}. 
We observed increasing the number of classes in the softmax layer led to a reduction in model accuracy and efficiency. We assume this decline likely stems from the model's lightweight design, limiting capture of global context \& long-term sequences.

\begin{table}[h!]
\centering
\caption{Comparison of the Proposed Model Against Existing Methods.}
\footnotesize
\begin{tabular}{|p{1cm}|p{2.3cm}|p{1.15cm}|p{1cm}|p{1.15cm}|}
\hline
\textbf{Study} & \textbf{Method} & \textbf{Dataset} & \textbf{Accuracy} & \textbf{Remarks} \\

\hline
~\cite{Agrawal2023ECGAuth} & 1D-CNN,LSTM & PTB & 93.6\%, 99.69\% & AUC 0.992  \\

\hline
~\cite{Hemrajani2023} & MobileNetV1 \newline MobileNetV1+LSTM \newline MobileNetV1+GRU & PhysioNet Apnea-ECG dataset  & 89.5\% \newline 90.0\% \newline 90.29\% & only detect sleep apnea \\ 

\hline
~\cite{Wang2024} & 1D (CNN) and self-supervised contrastive learning& PTB\newline MIT-BIH \newline ECG-ID & 99.15\% \newline 98.67\%  \newline 98.77\% 
& AUC 0.95-0.99 \\

\hline
~\cite{Safie2024} & Wigner-Ville distribution and GoogleNet architecture & NSRDB\newline MITDB & 99.004\% \newline 99.3\%
& NSRDB 0.8 \newline MITDB 1.13\\

\hline
\cite{BicakciYesilkaya2025} & GoogleNet, ResNet50 and DenseNet201 & PhysioNet Vollmer Dataset & Best 97.4\% & EER 0.16\% - 30.48\% \\

\hline
\multirow{4}{*}{\textbf{Proposed}} & \multirow{4}{*}{\textbf{MobileNetV1+GRU}} & \textbf{ECG-ID} & \textbf{99.34\%} & \textbf{EER 0.0091} \\
\cline{3-5}
 &  & \textbf{MIT-BIH} & \textbf{99.31\%} & \textbf{EER 0.00013} \\
\cline{3-5}
 &  & \textbf{CYBHi} & \textbf{91.74\%} & \textbf{AUC-ROC 0.9985} \\
\cline{3-5}
 &  & \textbf{PTB} & \textbf{98.49\%} & \textbf{AUC-ROC 0.9999} \\
\hline
\end{tabular}
\label{tab:ecg_comparison}
\end{table}

\section{Conclusion}
In this paper, the proposed MobileNetV1+GRU model with CWT-transformed ECG scalograms achieved 91.74–99.34\% accuracy across four diverse datasets, demonstrating promising robustness and generalization under noisy conditions. Hybrid preprocessing (fiducial and non-fiducial) was applied, alongside FedAvg for privacy-preserving learning and FGSM adversarial attacks to assess model vulnerabilities. 
Our proposed hybrid model outperformed existing methods, achieving SOTA results \&  capable of ECG classification, cardiac monitoring, and heart condition detection. Future work focuses on transformer-based multi-factor authentication robust models with diverse physiological conditions, clinical trials, advanced automated preprocessing, and improved spoofing resistance.

\clearpage
\vfill\pagebreak

\end{document}